\documentclass[aps,prl,twocolumn,showpacs,superscriptaddress]{revtex4}
\usepackage{graphicx}
\usepackage{epsfig}
\usepackage{amsmath}
\usepackage{color}
\usepackage{here}

\begin{document}

\title{The Gouy phase shift in nonlinear interactions of waves}
\author{Nico Lastzka}
\affiliation{Institut f\"ur Gravitationsphysik,
Leibniz Universit\"at Hannover and Max-Planck-Institut f\"ur
Gravitationsphysik (Albert-Einstein-Institut), Callinstr. 38,
30167 Hannover, Germany}
\author{Roman Schnabel}
\affiliation{Institut f\"ur Gravitationsphysik,
Leibniz Universit\"at Hannover and Max-Planck-Institut f\"ur
Gravitationsphysik (Albert-Einstein-Institut), Callinstr. 38,
30167 Hannover, Germany}

\date{\today}

\begin{abstract}
We theoretically analyze the influence of the Gouy phase shift
on the nonlinear interaction between waves of  different  frequencies.
We focus on $\chi^{(2)}$ interaction of optical fields, e.g. through birefringent crystals, and
show that focussing, stronger than suggested by the Boyd-Kleinman factor, can further improve nonlinear processes. An increased value of 3.32 for the optimal focussing parameter for a single pass process is found. The new value builds on the compensation of the Gouy phase shift by a spatially varying, instead constant, wave vector phase mismatch.
We analyze the single-ended, singly resonant standing wave nonlinear cavity and show that in this case
the Gouy phase shift leads to an additional phase during backreflection.
Our numerical simulations may explain ill-understood experimental  observations in such devices.
\end{abstract}

\pacs{03.67.-a, 42.50.-p, 03.65.Ud}
\maketitle

\noindent Nonlinear interactions of waves, in particular those of
optical fields, have opened new research areas and have found various
applications. In general, waves of different frequencies are coupled
via nonlinear media, like birefringent crystals. Examples are the
production of higher harmonics of laser radiation \cite{FHPW61}, the
generation of tunable frequencies through optical parametric
oscillation \cite{GMi65} and the generation of nonclassical light
\cite{WKHW86,GMa87} for high precision metrology \cite{Cav81,VCHFDS06},
fundamental tests of quantum mechanics and quantum information
\cite{YHa86}.
The efficiency of a nonlinear process depends on parameters of the nonlinear medium, and generally increases
with higher intensities  of the fields involved and with better phase matching of their wave fronts. To
achieve strong nonlinear interactions, pulsed laser radiation, strong focussing and, especially for
continuous wave radiation, intensity build-up in resonators are used. In plane wave theory perfect phase
matching is achieved if the wave fronts of interacting fields propagate with the same velocity. For focussed
laser beams, however, this is not true because of the well-known Gouy phase shift. This phase shift occurs
due to the spatial confinement of a focussed wave and generally depends on the spatial mode as well as the
frequency of the wave \cite{Feng2001}. The influence of focussing into a nonlinear medium has been
investigated by Boyd and Kleinman in great detail \cite{Boyd1968}. They discovered that the efficiency of the
nonlinear process does not monotonically increase with decreasing focal size. They especially considered the
lowest order nonlinearity that enables second harmonic generation (SHG) and optical parametric amplification (OPA) and is described by the susceptibility $\chi^{(2)}$, and numerically found an optimum
factor between the length of the nonlinear crystal and the Rayleigh range of the focussed Gaussian beam for a
single pass through the crystal.

In this Letter we show that focussing stronger than suggested by the
Boyd Kleinman factor can further improve nonlinear processes. We show that
this effect can be understood by considering the Gouy phase shift
between the interacting waves. We also show that the Gouy phase
shift results in a non-trivial phase mismatch problem in 
standing wave cavities. 

Boyd and Kleinman have found that the maximum nonlinear coupling between two Gaussian beams of fundamental
(subscript 1) and second harmonic waves (subscript 2)  is achieved for a positive wave vector phase mismatch
$\Delta k\!=\!2|k_{1}|\!-\!|k_{2}|\! >\! 0$ which increases with decreasing waist size of the beam. For a
single pass through a nonlinear crystal of length $L$ they numerically found the optimal focussing parameter
given by the relation
\begin{equation}
\xi := \frac{L}{2z_\textnormal{R}} = 2{.}84\,,
\label{FocusingParameter}
\end{equation}
where $z_\textnormal{R}=\pi w_0^2 n/\lambda$ is the Rayleigh range of the
beams inside the crystal, and $w_0$, $n$ and $\lambda$ are the beam's waist size, refractive index and wavelength, respectively.
We first show that the Boyd-Kleinman factor according
to Eq.\,(\ref{FocusingParameter}) is a consequence of maximizing the
intensity of the mean pump field inside the nonlinear medium under the constraint of the Gouy phase shift. In a $\chi^{(2)}$ medium the
nonlinear interaction is described by the following set of differential equations
\begin{align}
  \partial_z E_{0,1}(z) &\propto E_{0,1}^*(z)E_{0,2}(z)\cdot g^*(z)\,,\label{deq1} \\
  \partial_z E_{0,2}(z) &\propto E_{0,1}^2(z)\cdot g(z)\,,\label{deq2}
\end{align}
\begin{equation}
  g(z) := \frac{e^{i\Delta kz}}{1+i\frac{z-z_0}{z_\textnormal{R}}}
  = \frac{w_0}{w(z)} e^{i(\Delta kz+\Delta\phi(z))}\,,\label{g(z)}
\end{equation}
where $E_{0,1}$, $E_{0,2}$ are the electrical fields of the fundamental and
the harmonic mode in the focal center at  position $z_0$, and $\Delta
k=4\pi/\lambda\,\Delta n$ is the phase mismatch between the two
interacting modes. Here we use the following abbreviations
\begin{align}
  \Delta\phi(z) &= -\textnormal{arctan}\left(\frac{z-z_0}{z_\textnormal{R}}\right)\,,\label{gouy}\\
  w(z) &= w_0\sqrt{1+\left(\frac{z-z_0}{z_\textnormal{R}}\right)^2}\,,
\end{align}
where $w(z)$ corresponds to the beam width at the position $z$.
In plane wave theory one finds $g(z)=\exp(i\Delta kz)$ and
$\Delta\phi=0$, and equal indices of refraction for the two
interacting modes provide the maximum nonlinearity. When focussing the
beam into a nonlinear material however, there is a non zero phase
difference $\Delta\phi$. Since the phase difference between a plane
wave and a focussed Gaussian beam is given by the Gouy phase shift,
$\Delta\phi$ should be the difference of two such phase
shifts. When considering phase shifts between different oscillator
frequencies, phases have to be frequency normalized.
We therefore introduce the Gouy phase shift  normalized to the optical frequency of mode $i$
\begin{equation}
\tilde\phi_G(\omega_i) :=\frac{\phi_G}{\omega_i}=
-\frac{(m+n+1)}{\omega_i}\,\textnormal{arctan}\left(\frac{z-z_0}{z_\textnormal{R}}\right)\,,
\end{equation}
where m and n describe the spatial Hermite-Gaussian modes (TEM$_{mn}$).
If we now normalize $\Delta\phi$ to the harmonic
frequency we find
\begin{equation}
  \frac{\Delta\phi}{\omega_2} =
  \tilde\phi_G(\omega_1)-\tilde\phi_G(\omega_2)\,.
\end{equation}
We point out, that for an optimized nonlinear interaction of Gaussian beams the Rayleigh ranges are identical for all modes involved. In the case of frequency conversion of a single pump field this is automatically realized by the nonlinear process. For the $\chi^{(2)}$-processes considered in
Eqs.\,(\ref{deq1})-(\ref{g(z)}) ($m=n=0$) we find
\begin{equation}
  \Delta\phi = \phi_G \,.
\end{equation}

\begin{figure}[t]
\centerline{\includegraphics[width=8.3cm]{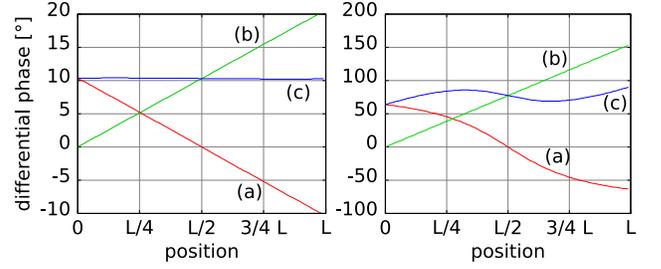}}
\caption{(Color online) Left: For weak focussing into the nonlinear medium
 ($\xi=0.18$) the Gouy phase shift can be compensated by choosing
 $\Delta k=1/z_\textnormal{R}$, and perfect  phase matching can be realized
 over the full crystal length. Right: For stronger focussing
 ($\xi=2.03$) a constant $\Delta k$ can not provide perfect phase
 matching. (a) Gouy phase shift $\Delta\phi$, (b) compensating phase
 $\Delta kz$, (c) overall phase $\phi_0$, where a constant value
 describes perfect phase matching.}
 \label{gouy_sp}
\end{figure}

From this one can conclude that the Gouy phase shift leads to a nonperfect matching of the (nonplanar) phase
fronts in nonlinear processes. To quantify this effect we define the
\emph{effective nonlinearity} $\kappa$ of the process. This quantity is proportional to the conversion efficiency
in SHG as well as to the optical gain of OPA. For weak interaction, i.\,e. the pump field is not depleted by the nonlinear interaction,
$\kappa$ is given by
\begin{equation}
\kappa := \left|\int dz\,\frac{g(z)}{w_0}\right|^2\,,
\end{equation}
where the integration is taken over the whole interaction length.
For a single pass through a nonlinear medium of length $L$ the effective
nonlinearity is given by
\begin{equation}
  \kappa_{\textnormal{sp}} =\left|\int_0^L dz\,\frac{e^{i(\Delta kz
  +\phi_G(z))}}{w(z)}\right|^2\,.
\end{equation}

This quantity is maximized if the averaged field strength inside the crystal is maximized,
i.\,e. if the focus is placed in the crystal center and if the condition
\begin{equation}
  \Delta kz + \Delta\phi(z) = \phi_0 = \textnormal{const.}\label{pm}
\end{equation}
is satisfied. In this case all partial waves are produced exactly \emph{in} phase to each other, and perfect
phase matching is realized.

Curves (a) in Fig. \ref{gouy_sp} show the differential Gouy phase shifts $\Delta\phi(z)=\phi_G(z)$ for weak
and strong focussing, respectively. For weak focussing  the gouy phase shift evolves linearly inside the
medium, and one can compensate this phase mismatch by choosing $\Delta k=1/z_\textnormal{R}>0$, as found by Boyd and
Kleinman \cite{Boyd1968}. For stronger focussing, however, it is not possible to achieve perfect compensation
from $\Delta k$ that is constant over the crystal. Curves (b) show the compensating linear phase $\Delta kz$
that is due to the propagation inside the medium and curves (c) show the total phase $\phi_{0}$.
The value of $\Delta k$ was chosen to provide the lowest variance of $\phi_0$ over the whole interaction range.

\begin{figure}[t]
\centerline{\includegraphics[width=8.0cm]{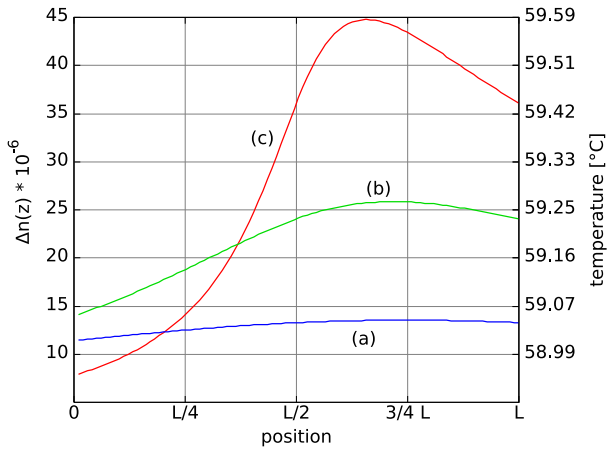}}
 \caption{(Color online) Change of refractive index along the crystal to compensate for the Gouy phase shift, for three different strengths of focussing. The temperature
 scale on the right corresponds to a MgO(7\%)\,:\,LiNbO$_3$ crystal
 that has been used in \cite{VCHFDS06}. (a) $\xi=0.55$, (b)
 $\xi=1.14$, (c) $\xi=3.32$.}
 \label{gouy_compensation}
\end{figure}

We now show that it is possible to realize perfect phase matching for an arbitrary focussing by
applying the following position dependent index of refraction
\begin{align}
  \Delta n_{\textnormal{sp}}(z)=\frac{\lambda}{4\pi}\Delta k(z) &=
  \frac{\lambda}{4\pi}\frac{\phi_0+\textnormal{arctan}(\frac{z-z_0}{z_\textnormal{R}})}{z}\,,
\end{align}
where the constant value of the phase $\phi_0$ is set by the Gouy phase at the entrance surface of the nonlinear medium
\begin{equation}
  \phi_0=\textnormal{arctan}\left(\frac{z_0}{z_\textnormal{R}}\right)=\phi_G(0)\, .
\end{equation}
Fig. \ref{gouy_compensation} shows $\Delta n_{\textnormal{sp}}(z)$ for a nonlinear crystal of length $L$ for different
focussing parameters $\xi$, with focus placed into the crystal's center. Curves (a) to (c) could experimentally be realized by applying an appropriate temperature gradient
along $z$-direction. Alternatively, an electrical field applied to the crystal could be used.
The temperature values on the right vertical axis constitute an example for 7\,\%
magnesium-oxide-doped lithium niobate (MgO:LiNbO$_3$). Data was deduced from measurements of the nonlinear efficiency of crystals used in type I OPA in \cite{VCHFDS06}.

The effective nonlinearity for the single pass setup with perfect Gouy phase compensation can be written as
\begin{equation}
  \kappa_{\textnormal{sp}}= \frac{1}{{\xi}}\ln^2\left(\frac{\sqrt{1+\xi^2}+\xi}
          {\sqrt{1+\xi^2}-\xi}\right)\,,
\end{equation}
with $\xi=L/2z_\textnormal{R}$. The numerical optimization of the above expression leads to
\begin{equation}
\xi_\textnormal{opt}=3{.}32\,.
\end{equation}
Our result means that with optimum, position dependent phase matching, the optimal waist size is approximatly
$7{.}5\,\%$ smaller than suggested by the Boyd-Kleinman factor, and according to this, the effective
nonlinearity is further increased by $4{.}4\,\%$.

We now analyze if the nonlinear interaction in standing-wave cavities can be similarily improved. Standing
wave cavities, in particular in the form of a singly-resonant, single-ended cavity, i.\,e. with one mirror of almost perfect
reflectivity, are frequently used in quantum and nonlinear
optics \cite{Paschotta1994,VCHFDS06} . In such cavities waves that
propagate in two different directions interfere with each other, and the differential phases introduced by
the reflections at the cavity mirrors have to be considered.
Paschotta \emph{et al.} have investigated the phase difference $\Delta\varphi$ introduced from back reflection and have suggested an appropriate design of the high reflectivity dielectric multi-layer
coating to annihilate any additional phase shift. It can be shown that the effective nonlinearity for a doublepass of \emph{plane}
waves through a crystal depends on the $\Delta\varphi$ in the following way
\begin{equation}
\kappa_{\textnormal{dp,pw}} = \frac{\sin^2\left(\frac{\Delta k L}{2}\right)}{\left(\frac{\Delta k L}{2}\right)^2}\cdot \cos^2\left(\frac{\Delta k L}{2}+\frac{\Delta\varphi}{2}\right)\,.
\label{doublepass}
\end{equation}
For the calculation of the doublepass effective nonlinearity in the case of focussed Gaussian beams we model the system with a nonlinear medium of length $2L$ with a thin lense at position $L$ that refocuses the beam.
In this way we obtain two waists at positions $z_0$ and $z_0'=2L-z_0$ of size $w_0$ indicating the way to the
endmirror and the way back, respectively. Now we integrate over $2L$ and find the following expression for
the effective nonlinearity for a double pass of the fundamental Hermite-Gauss mode
\begin{align}
  \kappa_{\textnormal{dp}} &= \frac{1}{w_0^2}\cdot\left|\int_0^L
  dz\,g(z,z_0)+\int_L^{2L} dz\,g(z,z_0')e^{i\Delta\varphi}\right|^2\nonumber\\
  &= \frac{1}{w_0^2}\cdot\left|\int_0^L
  dz\,\left[g(z,z_0)+g(z+L,z_0')e^{i\Delta\varphi}\right]\right|^2\nonumber\\
  &= \left|\int_0^L
 dz\,\frac{e^{i(\Delta kz+\phi_G(z))}}{w(z)}\right.\times\nonumber\\
 &\phantom{=|\int_0^L}
  \left.\left(1+\frac{w(z)}{w'(z)}e^{i(\phi_G'(z)-\phi_G(z)+\Delta
  kL+\Delta\varphi)}\right)\right|^2\,,\label{kappaDP}
\end{align}
where $w'(z)$ and $\phi_G'(z)$ belong to the focus at position $z_0'$. $\Delta\varphi$ is again the differential phase that may be introduced by the coating of the back reflecting mirror.
%
We first consider the special case of weak focussing, i.~e. $|z-z_0|/z_\textnormal{R}\ll 1\quad\forall\,z\in [0,\,L]$, and simplify the above expression as follows
\begin{equation}
  \kappa_{\textnormal{dp}} = \frac{\sin^2\left(\frac{\Delta
  k'L}{2}\right)}{\left(\frac{\Delta
  k'L}{2}\right)^2}\cdot\cos^2\left(\frac{\Delta
  k'L}{2}+\frac{\Delta\varphi'}{2}\right)\,,
\end{equation}
where $\Delta k' := \Delta k-1/z_\textnormal{R}$ and
\begin{equation}
  \Delta\varphi' := \Delta\varphi+2(L-z_0)/z_\textnormal{R}\,.\label{varphi}
\end{equation}
This expression has the same form as the one for plane waves as given in Eq.\,(\ref{doublepass}). However, an additional phase shift appears. This phase shift is a result of spatial confinement and the swapping in sign of the wave front's radius of curvature during reflection, and corresponds to minus
twice the Gouy phase in the limit considered here.
From the expression of $\Delta\varphi'$ in Eq.\,(\ref{varphi}) it follows that this additional phase jump vanishes if the waist is located exactly at the back reflecting surface. In this case we have plane wave fronts at the end mirror and therefore the system is similar to a single pass through a nonlinear medium of length $2L$.

\begin{figure}[t]
\centerline{\includegraphics[width=8.0cm]{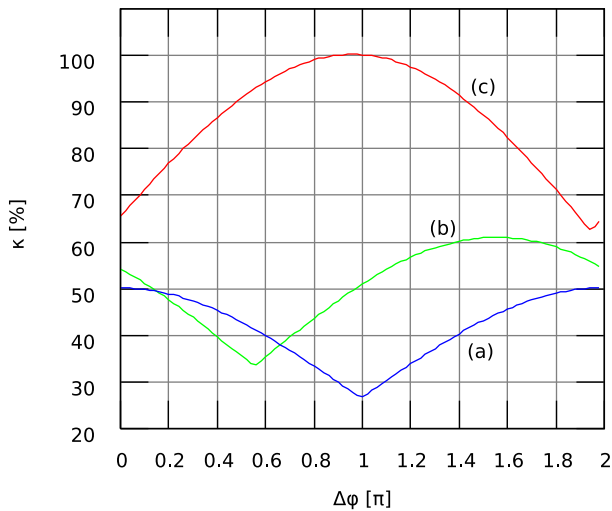}}
 \caption{
(Color online) Effective nonlinearity $\kappa_{\textnormal{dp}}$ (normalized) versus the differential phase
introduced by the back reflecting surface $\Delta\varphi$. For Gaussian beams with waist position on the
reflecting surface, $\Delta\varphi=0$ generally provides the highest $\kappa_\textnormal{dp}$, (a) (here with
strong focussing $\xi=2.84$, $z_0=L$); the same result is found for plane waves. Contrary, for $\xi=2.84$ and
$z_0=L/2$ we find  $\Delta\varphi=\pi$ (c). The result for weaker focussing is shown in (b) ($\xi=0.775$,
$z_0=L/2$). }
 \label{hrPhase}
\end{figure}

We now examine $\kappa_{\textnormal{dp}}$ for reflected Gaussian beams of arbitrary focussing parameters
$\xi$ and the waist located in the center of the medium. Eq.\,(\ref{kappaDP}) yields
\begin{equation}
\kappa_{\textnormal{dp}} = 4\cos^2\!\!\left(\!\frac{\Delta kL+\Delta\varphi}{2}\!\right)\left|\int_0^L
\!\!dz\,\frac{e^{i(\Delta kz+\phi_G(z))}}{w(z)}\right|^2\!.
\end{equation}
The first term provides the optimal phase of the end mirror of $\Delta\varphi=-\Delta kL$. In turn
$\Delta k$ is again found by minimizing the variance of the term $\Delta kz+\phi_G(z)$. We obtain the following expression for the optimal differential phase $\Delta\varphi$ of
the end mirror versus focussing parameter
\begin{equation}
\Delta\varphi = -\frac{3}{\xi^2}\left[(1+\xi^2)\cdot\textnormal{arctan}(\xi)-\xi\right]\,.
\end{equation}
Fig.\,\ref{hrPhase} shows the effective nonlinearity versus $\Delta\varphi$ for three different standing wave cavity arrangements. In all cases the second harmonic wave is not resonant but simply back reflected. Curves\,(a) and (c) use the focussing parameter $\xi=2.84$. This value optimizes the effective nonlinearity of the cavity if the refractive index of the medium does not depend on direction of propagation of waves. This is evident from Fig.\,\ref{gouy_compensation} because the position dependent refractive indices are not symmetric with respect to the focal position and the back reflected wave would require different values. However, if one transfers the results from a single pass through the medium and realizes the required propagation direction dependent refractive index the optimum focussing parameter is again $\xi=3.32$.
Curve\,(a) represents  the case for focussing directly onto the back reflecting surface. Curve\,(c) shows the
effective nonlinearity for a waist position at the crystal's centre. In the latter case the best choice of
the differential phase at the back reflecting surface is $\Delta\varphi\approx\pi$. This is exactly the
opposite of what one might expect from plane wave theory, where the optimum phase is $\Delta\varphi=0$,
similar to curve (a).
Trace $(b)$ shows the effective nonlinearity for the focussing parameter that was chosen by Paschotta
\emph{et.\,al.} \cite{Paschotta1993}. In that paper a full
\emph{quantitative} comparison between experiment and theory of the nonlinearity in standing
wave cavities was conducted. In
their experiment the back reflecting mirror was designed to prevent a differential phase shift between the
two interacting modes and a value of $\Delta\varphi=0$ was chosen. However, their experimental data revealed
the effective nonlinearity to be $\approx 10\,\%$ smaller than expected from their calculations. From our
calculation it follows that the optimum phase for their setup was $\Delta\varphi\approx 1{.}55\,\pi$ and that
the chosen value of $\Delta\varphi=0$ decreased the effective
nonlinearity to about $90\,\%$ of the maximum
value. Our considerations are therefore in excellent agreement with experimental results given in that paper
and can solve the observed discrepancy.

In conclusion we have shown how for focussed waves the Gouy phase shift produces nonideal phase matching in
case of $\Delta k = 0$. For a single pass through a nonlinear medium the optimum focussing parameter is found
to be $\xi=3.32$. In this case a position dependent refractive index is required to further improve the
effective nonlinearity by $4{.}4\,\%$. For a double pass and for cavities an optimum focussing parameter
above $\xi=2.84$ can only be achieved with a refractive index that also depends on propagation direction. We
have also shown that the Gouy phase shift effects the optimum value for the phases introduced by cavity
mirrors, with a significant effect on the achievable effective nonlinearity. Our theoretical analysis shows
exact agreement with experimental data published elsewhere, and may lead to improved quantitative
descriptions of nonlinear cavities.



\begin{thebibliography}{99}

\bibitem{FHPW61} P. A. Franken, A. E. Hill, C. W. Peters, and G. Weinreich, Phys. Rev. Lett. \textbf{7}, 118
(1961).

\bibitem{GMi65} J. A. Giordmaine and R. C. Miller, Phys. Rev. Lett. \textbf{14}, 973 (1965).

\bibitem{WKHW86} L. Wu, H. J. Kimble, J. L. Hall, and H. Wu,  Phys. Rev. Lett. \textbf{57}, 2520 (1986).

\bibitem{GMa87} R. Ghosh and L. Mandel, Phys. Rev. Lett. \textbf{59}, 1903 (1987).

\bibitem{Cav81} C.~M.~Caves, Phys. Rev. D {\bf 23}, 1693 (1981).

\bibitem{VCHFDS06} H. Vahlbruch, S. Chelkowski, B. Hage, A. Franzen, K. Danzmann, and R. Schnabel, Phys. Rev. Lett. {\bf 97}, 011101 (2006).

\bibitem{YHa86} Y. Yamamoto and H. A. Haus, Rev. Mod. Phys. \textbf{58}, 1001 (1986).



\bibitem{Feng2001}
S. Feng and H.~G. Winful, Opt. Lett. {\bf 26,} 8 (2001)

\bibitem{Boyd1968} G.~D. Boyd and D.~A. Kleinman, Journal of Applied
  Physics {\bf 39,} 8 (1968).

\bibitem{Paschotta1994} R. Paschotta, K. Fiedler, P. K{\"u}rz, R. Henking, S. Schiller and J. Mlynek, Opt. Lett. {\bf 19},
  17 (1994).

\bibitem{Paschotta1993} R. Paschotta, K. Fiedler, P. K{\"u}rz and J. Mlynek, App. Phys. B {\bf 58},
  117 (1994).

%
%
%
\end{thebibliography}
\end{document}